# Quadrupling the stored charge by extending the accessible density of states

Mengyu Yan[1,2,†], Peiyao Wang[3,†], Xuelei Pan[1,†], Qiulong Wei[1], Jefferson Z. Liu[3,*], Yunlong Zhao[1], Kangning Zhao[1], Bruce Dunn[4,*], Jun Liu[2], Jihui Yang[2,*], Liqiang Mai[1,5,*]

[1]State Key Laboratory of Advanced Technology for Materials Synthesis and Processing, Wuhan University of Technology, Wuhan 430070, China.

[2]Department of Materials Science and Engineering, University of Washington, Seattle, Washington 98105, USA.

[3]Department of Mechanical Engineering, University of Melbourne, Victoria 3010, Australia.

[4]Department of Chemistry and Biochemistry, University of California, Los Angeles, California 90095, USA.

[5]Foshan Xianhu Laboratory of the Advanced Energy Science and Technology Guangdong Laboratory, Xianhu hydrogen Valley, Foshan 528200, China

*Correspondence to:

mlq518@whut.edu.cn, jihuiy@uw.edu, zhe.liu@unimelb.edu.au, bdunn@ucla.edu

† These authors contributed equally to this work

**Highlights**

- A nearly three-fold enhanced stored charge under the field effect in nanoscale electrochemical devices.

- This unusual increase in energy storage is attributed to the gradient fermi-level structure and the extended accessible density of states.

- This novel application of the field-effect in energy storage devices is applicable to the $MnO_2$, $MoS_2$, and the other redox-based energy storage materials.

**Context & Scale**

Nanosized energy storage devices play an essential role in maintenance-free microprocessors and nanoscale energy harvesting/sensing technologies. In this study, we proved the field-effect transistor, which is the core component in microprocessors, can also serve as an amplifier for the nanosized energy storage devices. A three-fold enhanced stored charge is achieved under the field effect, paving the way for the high performance nanosized self-power system.

**Summary**

Nanosized energy storage, energy-harvesting, and functional devices are the three key components for integrated self-power systems. Here, we report on nanoscale electrochemical devices with a nearly three-fold enhanced stored charge under the field effect. We demonstrated the field-effect transistor can not only work as a functional component in nanodevices but also serve as an amplifier for the nanosized energy storage blocks. This unusual increase in energy storage is attributed to having a wide range of accessible electronic density of states (EDOS), hence redox reactions are occurring within the nanowire and not being confined to the surface. Initial results with $MoS_2$ suggest that this field effect modulated energy storage mechanism may also apply to many other redox-active materials. Our work demonstrates the novel application of the field-effect in energy storage devices as a universal strategy to improve ion diffusion and the surface-active ion concentration of the active material, which can greatly enhance the charge storage ability of nanoscale devices. The fabrication process of the field-effect energy storage device is also compatible with microtechnology and can be integrated into other microdevices and nanodevices.

**Graphical Abstract**

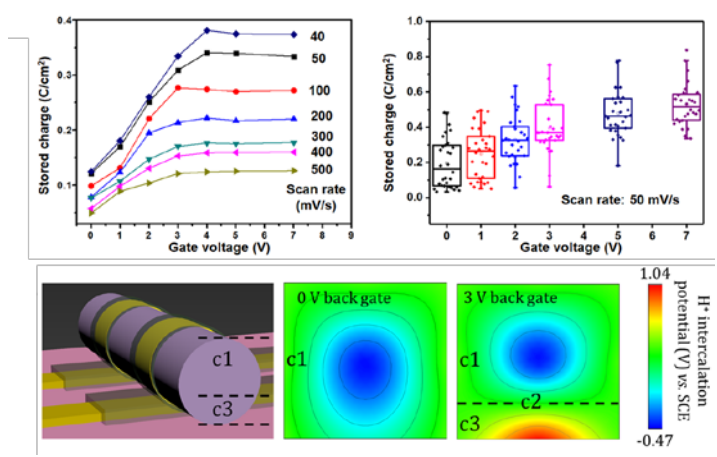

**Introduction**

Nanoscale devices, such as intracellular recording sensors, electron devices, solar cells, thermoelectric conversion devices, and nanogenerators[1, 2], show promise as fundamental building blocks in maintenance-free implantable biosensors[3-5]. Traditionally, an integrated device contains three key components: energy-harvesting, energy-storage, and functional components (e.g., sensors, data transmitters, and data controllers). An energy storage device can store discrete energy resources harvested by a solar cell, nanogenerator, or thermoelectric conversion device and then power the sensor and electronic device continuously. Thus, the development of nanoscale energy storage devices is meaningful and important as an indispensable component in integrated maintenance-free devices.

Field effects are widely used for nanoscale electronic devices, such as in transistors for logic circuits and amplifiers [6-10], in field-induced electronic phase transitions [11, 12], and in the modulation of the liquid/solid interface for liquid-gated interfacial superconductivity [13, 14]. Although the use of magnetic field in conjunction with electrochemical energy storage devices has shown some intriguing effects, it necessities magnetic constituents in the devices, hence, limiting its application.[15-17] To the best of our knowledge, the electrical field effects have not been integrated into any electrochemical energy storage device. Although electrochemical energy storage devices involve a change in electrical properties when charge transfer occurs [18, 19], it is highly interesting to explore whether modulation of EDOS is able to influence the energy storage processes *via* the induced gate-voltages. Thus, the overarching objective in the present paper is to determine whether such electric field effects can be used to influence and perhaps enhance the energy storage properties of an electrochemical device.

To explore the question of whether EDOS influence energy storage, we have designed a nanowire-based, field-effect energy storage device (FE-ESD, Figure 1a). The device is based on using a single manganese dioxide (α-MnO$_2$) nanowire of 30 nm in diameter and approximately 20 μm in length. The detailed fabrication procedure is shown in Figure S1. From the electrochemical point of view, the α-MnO$_2$ nanowire functions as the cathode while the large gold pad as the anode, with an electrolyte of 6 mol/L KOH. In the 'field-effect' part of the FE-ESD, an additional electrical potential (gate-voltage) is induced via Si *vs.* the ground. The highly doped Si substrate and the 300 nm SiO$_2$ layer serve as the back gate and dielectric layer, respectively (Figure 1b). The α-MnO$_2$ nanowire acts as a semiconductor channel in which the gate electrode is expected to influence the flow of charge. Thus, the FE-ESD integrates an electrochemical device structure with a gate electrode configuration while the α-MnO$_2$ nanowire serves as both the cathode electrochemically and as a conductive channel 'field-effect' wise. With this configuration, we are able to determine how the gate-voltage influences the electrode/electrolyte interface and the resulting energy storage behavior. The optical image (Figure 1c) shows that the α-MnO$_2$ nanowire is anchored by Au electrodes and that these electrodes are fully covered by an SU-8 passivation layer to avoid current leakage. We have tested the leakage currents of these nanowire devices from two sources, from the SiO$_2$ dielectric layer and from the current collector with SU-8 passivation layer to the electrolyte. The first one was determined to be less than 0.15 nA under 1 V (Figure S2c). The second was characterized (Figure S2d) using a 3-electrode cell in which the α-MnO$_2$ was charged to 0.4V *vs.* SCE, and the self-discharge properties were measured. A leakage current of 5 pA was obtained.

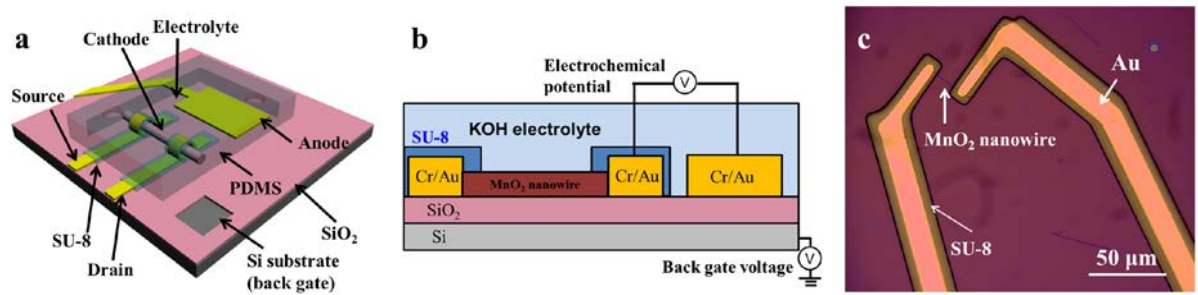

**Figure 1. Configuration of the field-effect energy storage device (FE-ESD). a, b.** a three-dimensional view and a cross-section of the FE-ESD based on a single α-$MnO_2$ nanowire. **c.** Optical image for the FE-ESD showing the α-$MnO_2$ nanowire in contact with two gold electrodes. SU-8 is used to cover the gold electrodes to prevent current leakage.

**Results and Discussion**

We have investigated the electrochemical performance of the FE-ESD at different gate-voltages by using a probe station, an electrochemical workstation, and a semiconductor device analyzer. Cyclic voltammetry (CV) was used to characterize the energy storage properties of devices. By comparing the CV of a FE-ESD with and without the α-$MnO_2$ (current flow through the α-$MnO_2$ and electrolyte, respectively), we are able to determine how much stored charge is attributable to the α-$MnO_2$ nanowire and the background. As shown in Figure S3, at a scan rate of 100 mV/s, the stored charge of the α-$MnO_2$ single-nanowire FE-ESD is almost $10^5$ times greater than that of the background. We subsequently applied the gate-voltage to the α-$MnO_2$ FE-ESD to determine its influence on energy storage. Upon increasing the gate-voltage from 0 to 3 V, the area under the CV curve at 100 mV/s is nearly tripled (Figure 2a). Specifically, the stored charge increases monotonically from ~100 mC/$cm^2$ at 0 V to 277 mC/$cm^2$ with a gate-voltage of 3 V. The scan rate dependence of the stored charge as a function of the gate-voltage is shown in Figure 2b. Even at a relatively high scan rate of 500 mV/s, the stored charge (120 mC/cm$^2$) at 3 V is still 2.5 times greater. Increasing the gate-voltage beyond 3 V does not provide any additional benefit as the stored charge remains constant in the range between 3 and 7 V (Figure 2c and S4A).

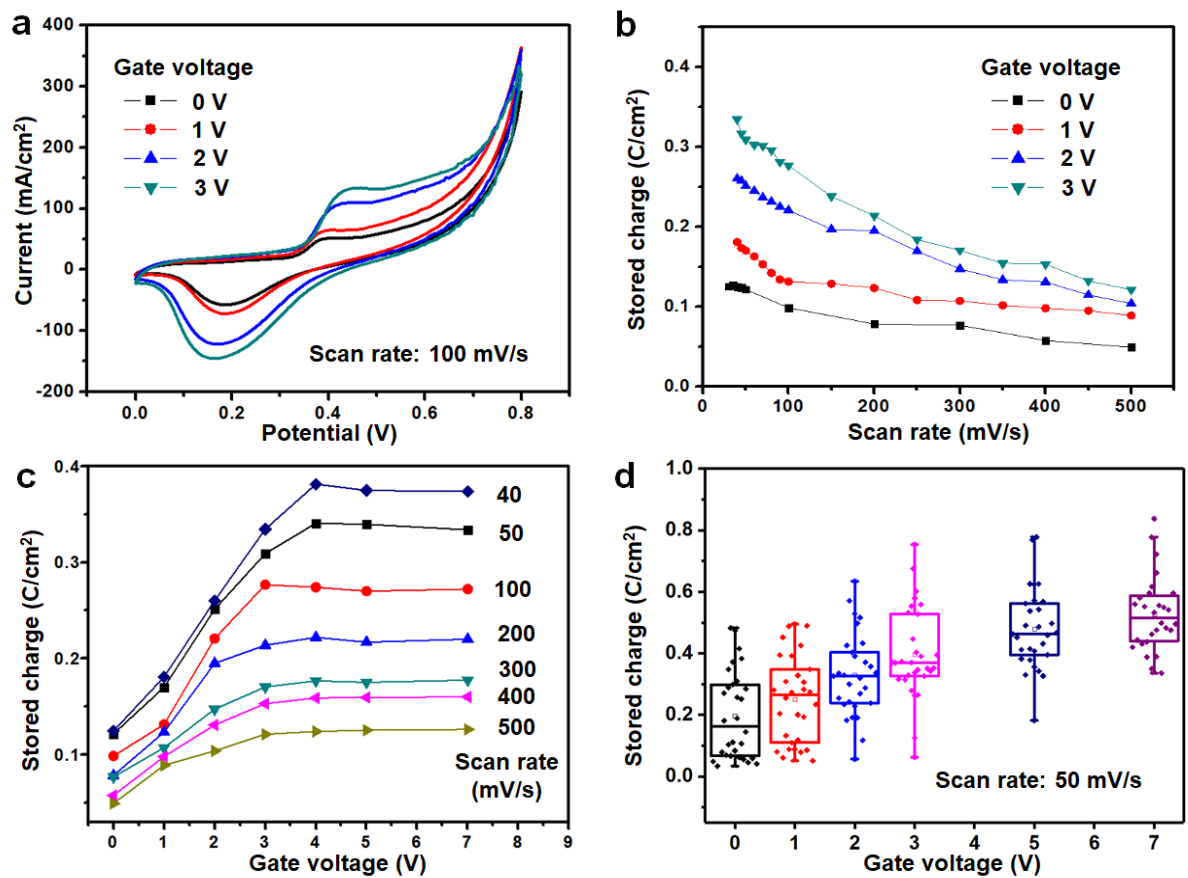

**Figure 2. Electrochemical performance of the α-MnO$_2$ single-nanowire FE-ESD. a.** CV curves at 100 mV/s with the gate-voltages of 0, 1, 2, and 3 V. **b.** The stored charge as a function of scan rate (30–500 mV/s). **c.** The stored charge as a function of the gate-voltage (0–7 V). **d.** The stored charge for thirty-one devices as a function of the gate-voltage at a scan rate of 50 mV/s. The highest and lowest horizontal lines in the boxes represent the upper and lower quartiles, respectively, whereas the middle line represents the mean.

We carried out two other series of experiment to verify the reproducibility of these gate-voltage results. In one case, 31 α-MnO$_2$ FE-ESDs were fabricated and tested at a scan rate of 50 mV/s. The data shown in Figure 2d indicate that although there are device-to-device variations, the stored charge at a gate-voltage of 3 V is, on average, two times greater than that at 0 V. We also carried out electrochemical three-electrode measurements as a reference electrode (saturated calomel electrode, SCE) was integrated into the device structure shown in Figure 1. As shown in Supplementary Figures 4b-d, similar conclusion as the two-electrode experiment can be drawn with respect to the gate-voltage and scan rate dependences. The three-electrode experiment (Figure S4b) also shows that oxidation and reduction peak occurs at 0.18 and 0.02 V *vs.* SCE, respectively. These peaks are attributed to the equilibrium potential of the Mn$^{4+}$/Mn$^{3+}$ redox [20-22].

Accompanying the stored charge increase with gate-voltage is the increased voltage offset between oxidation and reduction peaks (Figure S5). Usually, an increase in polarization indicates a decrease in the ion/electron diffusion rate and poor electrochemical performance. In our study, however, the stored charge increases as the voltage separation increases. To discern the influence of the gate-voltage, we studied the charge storage mechanism of the FE-ESD in greater detail. We assumed that the peak current (*i*) of the CV obeys a power-law relationship with the scan rate *v*

$$i = av^b, \qquad (1)$$

where *a* and *b* are adjustable values [23, 24]. The value of *b* provides insight into the charge storage mechanism as to whether the currents are limited by semi-infinite diffusion (*b* = 0.5) or surface-controlled (*b* = 1). Figure 3a shows a plot of log (*i*) of the cathodic peak *vs.* log (*v*) at different gate-voltages. For scan rates between 50 and 200 mV/s, *b* is 0.91 at a gate-voltage of 0 V, indicating that the kinetics are surface-controlled. This behavior is expected because of the α-MnO$_2$ nanowire geometry. As the gate-voltage increases to 3 V, *b* decreases continuously from 0.91 to 0.67. This suggests that the redox reaction is not only occurring at the surface of the nanowire but, because of a greater diffusion contribution, also extending further into the nanowire [25]. Thus, upon increasing the gate-voltage, there is a greater amount of energy storage as the redox reactions are not confined to the surface but penetrate into the 30 nm diameter wire.

We also use the CV results to quantify the stored charge associated with diffusion- and surface-controlled mechanisms. In this case, we use the relation

$$i(V) = a_1 v + a_2 v^{1/2},  \qquad (2)$$

where the current response (*i*) at a potential (*V*) can be described as a combination of two mechanisms, surface-controlled charge storage ($a_1 v$) and diffusion-controlled one ($a_2 v^{1/2}$) [26, 27]. We then calculate the values of $a_1$ and $a_2$ over the potential range (0 to 0.8 V) to obtain the amount of charge storage for each mechanism. At 50 mV/s, we find that the amount of stored charge associated with diffusion-controlled processes increases substantially from 0.076 C/cm$^2$ at 0 V to 0.242 C/cm$^2$ at 3 V. The increase in the surface-controlled contribution exhibits only a modest increase from 0.052 C/cm$^2$ to 0.097 C/cm$^2$. This analysis shows that the surface-controlled currents remain relatively constant with the application of the gate-voltage, but that the diffusion-controlled currents are largely responsible for the energy storage increase.

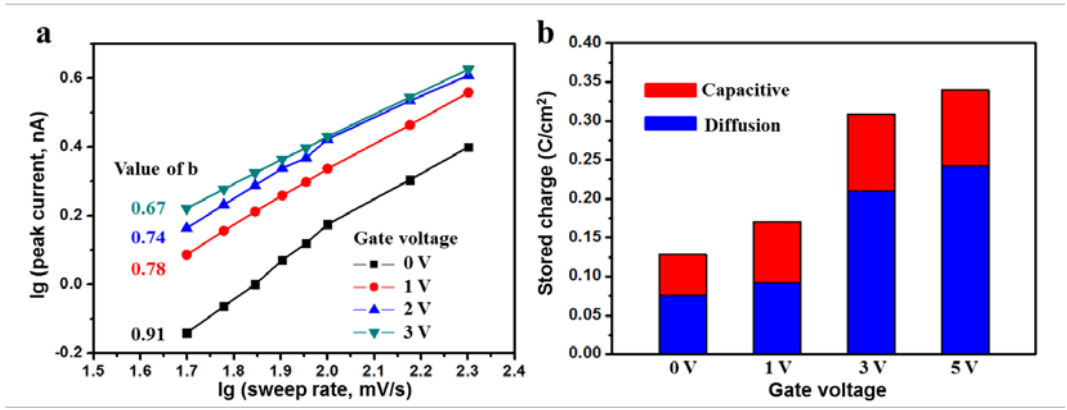

**Figure 3. Kinetics of the α-MnO$_2$ single-nanowire FE-ESD electrochemical behavior. a.** A plot of log(*i*) *vs.* log(*v*) for scan rates from 50 to 200 mV/s to determinate *b* in Equation (1). **b.** Bar chart showing the contributions of stored charge from capacitance and diffusion at a scan rate of 50 mV/s.

We further employed the open-circuit potential (OCP) to reveal the gate-voltage dependent Fermi-level of α-MnO$_2$ in the electrolyte. These Fermi-level positions, subsequently, were applied to the electronic band structures to understand the effects of the gate-voltage on the energy storage device performance. As shown in Figures 4a and b, the OCP of α-MnO$_2$ is -0.25 V *vs.* SCE when the gate-voltage is 0 V (equal to 4.38 eV *vs.* vacuum).  When the gate-voltage is increased to 3 V, the electrons are injected into the bottom-surface area of α-MnO$_2$, leading to a higher Fermi-level of 2.93 eV (-1.70 V *vs.* SCE; obtained from the OCP test); in the meantime, the Fermi-level in the α-MnO$_2$ upper-surface area remains unchanged (Figure 4b, c1 area).  To fully understand band-bending at the α-MnO$_2$-electrolyte interface, as well as how this influences the redox reaction, two parameters are introduced into the band-diagram. The first one is the Fermi-level in the 6M KOH aqueous solution, which is $\Phi_{SCE} - 0.059 \times pH = 3.57$ eV.  The other is the H$^+$ redox potential in α-MnO$_2$. In the band diagram, the incorporated ions are treated as the extrinsic impurities which occupy certain electronic levels within the α-MnO$_2$ bandgap. The H$^+$ energy level, locating at 4.84 eV (in 6M KOH electrolyte), is calculated according to Ref. 28.

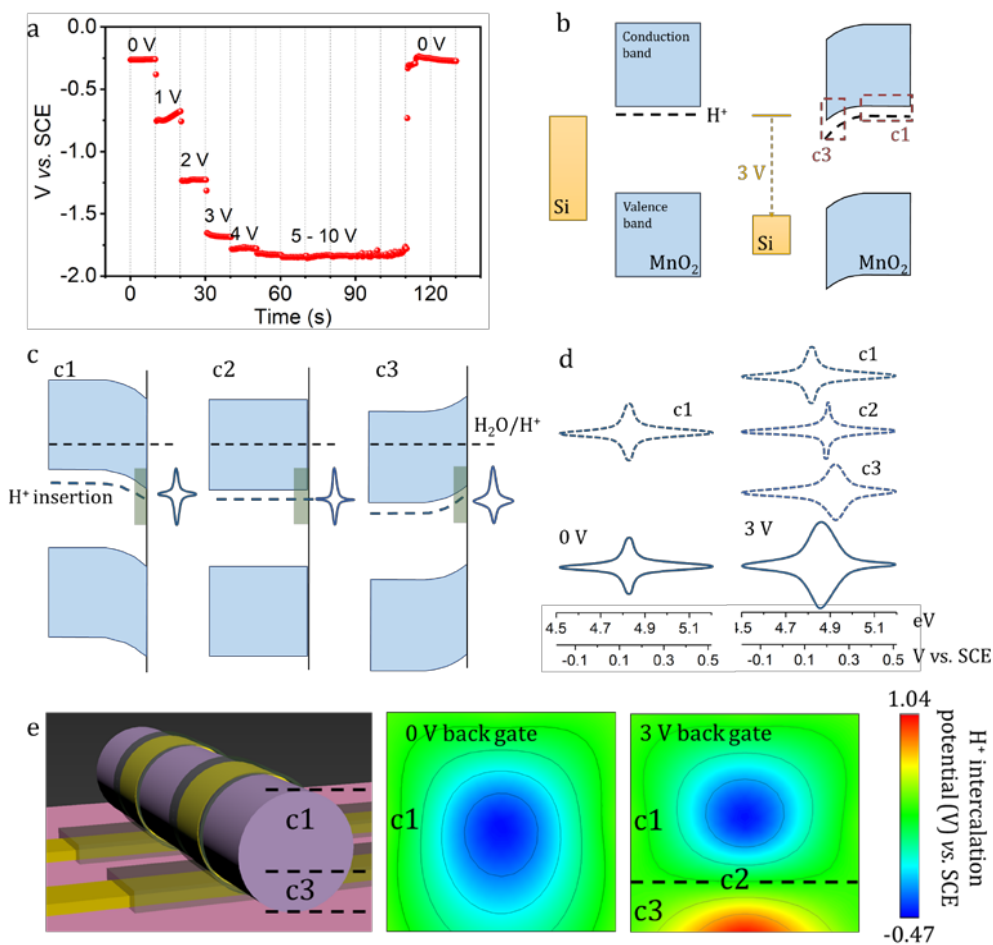

**Figure 4 Gate-voltage-dependent open-circuit potential and band-diagram of α-MnO$_2$ immersed in a 6M KOH aqueous solution. a.** Gate-voltage induced open-circuit potential changes in a 6 M KOH. **b.** (left figure) The band structure of the device without back gate-voltage. (right figure) The applied 3 V gate-voltage will move the conduction band, valence band, the Fermi-level, and H$^+$ energy level downward before the α-MnO$_2$ nanowire touching the electrolyte. **c. (c1)** The non-gate-controlled area, including the case without gate-voltage and the upper area of α-MnO$_2$ at 3 V back gate-voltage, will bend down; **(c2)** A specific state that no band-bending occurs. This phenomenon will likely appear in the middle part of α-MnO$_2$ at 3 V back gate-voltage; **(c3)** Opposite to the non-gate-controlled area shown in Figure 4c1, the bottom area of α-MnO$_2$ bends

up when in touch with the electrolyte. **d.** The simulated CV curve without back gate-voltage and at 3 V back gate-voltage. The band-bending as shown Figures 4c1-c3 leads to different CV curves. The reaction potential at 3 V back gate-voltage contains all situations shown in Figures 4c1, 4c2, and 4c3 resulting in broader redox peaks and upper-shift redox potential. **e.** The schematics of $H^+$ insertion potential in $MnO_2$. Under zero back-gate voltage, only the c1 state exists in $MnO_2$. For 3 V back-gate voltage, the higher $H^+$ insertion potential area emerges (c3), resulting in the higher redox potential.

In the electrochemical system, an electrochemical reaction should take place while the scanned potential windows by the electrolytes (eΦ) pass through the $H^+$ energy level as shown in Figure 4. When there is no gate-voltage, the electrochemical reaction should take place near eΦ = 4.84 eV. This value is reasonably consistent with our experimental result shown in Figure S4b, eΦ = 4.80 eV (0.12 V vs. SCE). The 0.04 eV difference may due to the downward band-bending at the α-$MnO_2$-electrolyte interface (Figure 4c1). As indicated in Figure 3a, the redox reactions are confined to the surface of the α-$MnO_2$ nanowire. Thus, the redox potential is only slightly influenced by the band-bending, decreasing from 4.84 to 4.80 eV.  Upon increasing the gate-voltage, the Fermi-level of α-$MnO_2$ moves up from 4.38 eV to 2.93 eV, lower than that of 6M KOH aqueous solution (3.57 eV), resulting in an upward band-bending (Figure 4c3). Therefore, higher redox potential is observed in Figure 2a.

Besides the higher redox potential, there are other two interesting phenomena that emerged by applying the back gate-voltages, the broader redox peaks and the greater stored charge (Figure 2a).

These two experimental results are explained in Figures 4d and 4e. With the gate-voltage, the energy state of α-MnO$_2$ bottom-surface is schematically shown in Figure 4c3, while its top-surface stay at the original state (shown in Figure 4c1). Hence there is a Fermi-level gradient in the α-MnO$_2$ nanowire along with the electric field direction. In the middle, there exists a point with no band-bending (Figure 4c2). The Fermi-level gradient increases the accessible EDOS participating the reaction as illustrated in Figure 4. As a result, CV at 3 V gate-voltage contains all of the three situations in Figures 4c1, 4c2, and 4c3. The reaction potential range (E) at 3 V gate-voltage is broader than that without gate-voltage, leading to the widening of redox-peaks. The H$^+$ insertion potential in MnO$_2$ is schematically shown in Figure 4e. A red (high potential) region emerges after applying the back-gate voltage, leading to a higher average redox penitential of the MnO$_2$ and the broadened redox peaks (as shown in Figure 2a). Furthermore, for a CV experiment, there is $t = E / v$, in which, $t$ is the reaction time for a specific redox reaction, $E$ the potential range of the redox reaction, and $v$ the scan rate. The reaction time $t$ will be longer after applying the gate-voltage, and thus more ions participating in the reaction.

In addition to the results for α-MnO$_2$, comparable experiments carried out using molybdenum disulfide (MoS$_2$) nanosheets show that the application of a gate-voltage leads to enhanced energy storage (Figure S6). Interestingly, graphite sheets show no improvement in charge storage, even at a high gate-voltage of 10 V (Figure S7). The reason for this is that the principal charge storage mechanism for graphene sheets, the electrochemical double-layer capacitance, has a minimal contribution from the redox reactions.

**Conclusions**

Our study shows that by modifying the configuration of a nanowire electrochemical cell to include a gate electrode, it is possible to increase the energy storage properties of the device. Analysis of the CV experiment shows that without the gate-voltage, redox reactions are confined to the surface of the α-$MnO_2$ nanowire; and that by increasing the gate-voltage, redox reactions extend beyond the nanowire surface and lead to the energy storage increase. The energy band diagram semi-quantitatively shows that the proton insertion energy level bending under gate-voltage contributes to the higher redox potential, greater energy storage, and the broader redox peaks. Finally, initial results for $MoS_2$ suggest that the gate-voltage induced increase in energy storage may be applicable to many other redox-active materials.

**Experimental Procedures**

**Synthesis of MnO$_2$ nanowires.** The α-MnO$_2$ nanowires were synthesized by a hydrothermal method following a previous report [29]. In a typical synthesis, 2 mmol of KMnO$_4$ and 2 mmol of NH$_4$F can be dissolved in 80 mL of water and stirred at 25 °C for 1 h. The solution is then poured into a 100 mL autoclave and heated at 180 °C for 48 h. After washing and drying, the α-MnO$_2$ nanowires, which are over 20 μm in length and 20 – 40 nm in diameter, are obtained. The large length and small diameter of our synthesize α-MnO$_2$ nanowires make it possible to fabricate single nanowire-based energy storage devices.

**Device fabrication and characterization.** A Cr/Au (5 nm/50 nm) counter electrode was first deposited onto a silicon wafer with a 300 nm thick layer of SiO$_2$ using the electron-beam lithography followed by physical vapor deposition and lift-off. The α-MnO$_2$ nanowire was immobilized by the Cr/Au (5 nm/150 nm) contact and a SU-8 passivation layer deposition. Finally, electrolyte (6 mol/L KOH) was used to cover the working electrode (α-MnO$_2$ nanowire) and counter electrode (Cr/Au pad) of the energy storage device. Cyclic voltammetry of the field-effect-modulated energy storage device and the rate behavior at different gate-voltages were measured using a probe station (Lake Shore PPT4), an electrochemical workstation (Autolab PGSTAT 302N), and a semiconductor device analyzer (Agilent B1500A).

**Acknowledgments**

This work was supported by the National Key Research and Development Program of China (2020YFA0715000, 2016YFA0202603), the National Natural Science Foundation of China (51872218, 51832004, 51521001), Foshan Xianhu Laboratory of the Advanced Energy Science and Technology Guangdong Laboratory (XHT2020-003). B.D. acknowledges support for this research from the Office of Naval Research.

**Author Contributions**

Liqiang Mai, Mengyu Yan, Jihui Yang conceptualized the project. Jihui Yang, Liqiang Mai, Bruce Dunn, and Jun Liu supervised this project. Mengyu Yan, Peiyao Wang, Xuelei Pan, Qiulong Wei, Yunlong Zhao performed the experiments. Jun Liu, Kangning Zhao, Mengyu Yan, Qiulong Wei, Liqiang Mai, Jihui Yang, Bruce Dunn discussed the results and commented on the manuscript.

**Declaration of Interests**

The authors declare that they have no competing financial interests.

**Supplementary Information**

Extended data figures and tables is available in the online version of the paper.

**References**

1. Hochbaum, A. I., Chen, R., Delgado, R. D., Liang, W., Garnett, E. C., Najarian, M., Majumdar, A., Yang, P. (2008). Enhanced thermoelectric performance of rough silicon nanowires. Nature *451*, 163-167.
2. Wu, W., Wang, L., Li, Y., Zhang, F., Lin, L., Niu, S., Chenet, D., Zhang, X., Hao, Y., Heinz, T. F. (2014). Piezoelectricity of single-atomic-layer $MoS_2$ for energy conversion and piezotronics. Nature *514*, 470-474.
3. Jiang, Z., Qing, Q., Xie, P., Gao, R., Lieber, C. M. (2012). Kinked p-n junction nanowire probes for high spatial resolution sensing and intracellular recording. Nano Letters *12*, 1711-1716.
4. Duan, X., Gao, R., Xie, P., Cohen-Karni, T., Qing, Q., Choe, H. S., Tian, B., Jiang, X.,

Lieber, C. M. (2012). Intracellular recordings of action potentials by an extracellular nanoscale field-effect transistor. Nature Nanotechnol *7*, 174-179.
5. Kucsko, G., Maurer, P., Yao, N. Y., Kubo, M., Noh, H., Lo, P., Park, H., Lukin, M. D. (2013). Nanometre-scale thermometry in a living cell. Nature *500*, 54-58.
6. Novoselov, K. S., Geim, A. K., Morozov, S., Jiang, D., Zhang, Y., Dubonos, S., Grigorieva, I., Firsov, A. (2004). Electric field effect in atomically thin carbon films. Science *306*, 666-669.
7. Radisavljevic, B., Radenovic, A., Brivio, J., Giacometti, V., Kis, A. (2011). Single-layer $MoS_2$ transistors. Nature Nanotechnol *6*, 147-150.
8. Britnell, L., Gorbachev, R., Jalil, R., Belle, B., Schedin, F., Mishchenko, A., Georgiou, T., Katsnelson, M., Eaves, L., Morozov, S. (2012). Field-effect tunneling transistor based on vertical graphene heterostructures. Science *335*, 947-950.
9. Huang, Y., Duan, X., Cui, Y., Lauhon, L. J., Kim, K.-H., Lieber, C. M. (2001). Logic gates and computation from assembled nanowire building blocks. Science *294*, 1313-1317.
10. Huang, Y., Duan, X., Wei, Q., Lieber, C. M. (2001). Directed assembly of one-dimensional nanostructures into functional networks. Science *291*, 630-633.
11. Dhoot, A. S., Yuen, J. D., Heeney, M., McCulloch, I., Moses, D., Heeger, A. J. (2006). Beyond the metal-insulator transition in polymer electrolyte gated polymer field-effect transistors. Proc. Natl. Acad. Sci. *103*, 11834-11837.
12. Dhoot, A. S., Israel, C., Moya, X., Mathur, N. D., Friend, R. H. (2009). Large electric field effect in electrolyte-gated manganites. Phys. Rev. Lett. *102*, 136402.
13. Ye, J., Inoue, S., Kobayashi, K., Kasahara, Y., Yuan, H., Shimotani, H., Iwasa, Y. (2010). Liquid-gated interface superconductivity on an atomically flat film. Nat. Mater. *9*, 125-128.
14. Bollinger, A. T., Dubuis, G., Yoon, J., Pavuna, D., Misewich, J., Božović, I. (2011). Superconductor-insulator transition in $La_{2-x}Sr_xCuO_4$ at the pair quantum resistance. Nature *472*, 458-460.
15. Zhu, J., Chen, M., Wei, H., Yerra, N., Haldolaarachchige, N., Luo, Z., Young, D. P., Ho, T. C., Wei, S., Guo, Z. (2014). Magnetocapacitance in magnetic microtubular carbon nanocomposites under external magnetic field. Nano Energy *6*, 180-192.
16. Wei, H., Gu, H., Guo, J., Wei, S., Liu, J., Guo, Z. (2013). Silica doped nanopolyaniline with endured electrochemical energy storage and the magnetic field effects. The Journal of Physical Chemistry C *117*, 13000-13010.
17. Zhu, J., Chen, M., Qu, H., Luo, Z., Wu, S., Colorado, H. A., Wei, S., Guo, Z. (2013). Magnetic field induced capacitance enhancement in graphene and magnetic graphene nanocomposites. Energy & Environmental Science *6*, 194-204.
18. Liu, X., Li, B., Li, X., Harutyunyan, A. R., Hone, J., Esposito, D. V. J. N. L. (2019). The Critical Role of Electrolyte Gating on the Hydrogen Evolution Performance of Monolayer $MoS_2$. Nano Letters.
19. Tatara, R., Karayaylali, P., Yu, Y., Zhang, Y., Giordano, L., Maglia, F., Jung, R., Schmidt, J. P., Lund, I., Shao-Horn, Y. J. J. o. T. E. S. (2019). The effect of electrode-electrolyte interface on the electrochemical impedance spectra for positive electrode in Li-ion battery. Journal of the Electrochemical Society *166*, A5090-A5098.
20. Kozawa, A., Powers, R. (1966). The Manganese Dioxide Electrode in Alkaline Electrolyte; The Electron-Proton Mechanism for the Discharge Process from $MnO_2$ to $MnO_{1.5}$. Journal of the Electrochemical Society *113*, 870-878.
21. Qu, D., Conway, B., Bai, L., Zhou, Y., Adams, W. (1993). Role of dissolution of Mn (iii) species in discharge and recharge of chemically-modified $MnO_2$ battery cathode materials. Journal of applied electrochemistry *23*, 693-706.
22. Fiedler, D. A. (1998). Rapid evaluation of the rechargeability of c-$MnO_2$ in alkaline media by abrasive stripping voltammetry. Journal of Solid State Electrochemistry *2*, 315-320.
23. Lindström, H., Södergren, S., Solbrand, A., Rensmo, H., Hjelm, J., Hagfeldt, A., Lindquist, S.-E. (1997). $Li^+$ ion insertion in $TiO_2$ (anatase). 2. Voltammetry on nanoporous films. J. Phys. Chem. B *101*, 7717-7722.
24. Choi, C., Ashby, D. S., Butts, D. M., DeBlock, R. H., Wei, Q., Lau, J., Dunn, B. J. N. R. M. (2019). Achieving high energy density and high power density with pseudocapacitive materials. Nature Reviews Materials, 1-15.
25. Brezesinski, T., Wang, J., Tolbert, S. H., Dunn, B. (2010). Ordered mesoporous α-$MoO_3$ with iso-oriented nanocrystalline walls for thin-film pseudocapacitors. Nat. Mater. *9*, 146-151.
26. Wang, J., Polleux, J., Lim, J., Dunn, B. (2007). Pseudocapacitive contributions to electrochemical energy storage in $TiO_2$ (anatase) nanoparticles. J. Phys. Chem. C *111*, 14925-14931.
27. Sathiya, M., Prakash, A., Ramesha, K., Tarascon, J. M., Shukla, A. (2011). $V_2O_5$-anchored carbon nanotubes for enhanced electrochemical energy storage. J. Am. Chem. Soc. *133*, 16291-16299.
28. Young, M. J., Holder, A. M., George, S. M., Musgrave, C. B. (2015). Charge storage in cation incorporated α-MnO2. Chemistry of Materials *27*, 1172-1180.
29. Hu, P., Yan, M., Wang, X., Han, C., He, L., Wei, X., Niu, C., Zhao, K., Tian, X., Wei, Q. (2016). Single-nanowire electrochemical probe detection for internally optimized mechanism of porous graphene in electrochemical devices. Nano Letters *16*, 1523-1529.